\documentclass{optica-article}

\journal{opticajournal} 

\articletype{Research Article}

\usepackage{lineno}

\usepackage{graphicx}
\usepackage{dcolumn}
\usepackage{bm}
\usepackage{float}
\usepackage{amsmath}
\usepackage{subcaption}

\begin{document}

\title{STUDY OF MAGNIFICATION AND ANGULAR RESOLUTION OF A SINGLE WATER DROPLET PLACED ON A GLASS SURFACE}%

\author{Luka Chakhnashvili,\authormark{1,*} Giorgi Bakhtadze,\authormark{2}}

\address{\authormark{1}9 Petre Chaiakovski St, I.Vekua School of Physics and Mathematics, 0114 Tbilisi, Georgia\\
\authormark{2}9 Petre Chaiakovski St, I.Vekua School of Physics and Mathematics, 0114 Tbilisi, Georgia\\}

\email{\authormark{*}chakhnashvililuka@gmail.com}


\begin{abstract}
   In this study, we investigate the magnification and angular resolution of a single water droplet positioned on a glass surface, functioning as an optical imaging system. Through theoretical analysis of the droplet's shape, magnification, and angular resolution, we derive predictions that are subsequently validated through experiments. Our study explores the impact of key parameters, including droplet size, the distance between the droplet and the object, and the contact angle, on the aforementioned optical characteristics. Our findings reveal that smaller droplets exhibit higher magnification at shorter object-to-droplet distances and demonstrate superior resolving capability (i.e., smaller angular resolution).
\end{abstract}

\section{Introduction}
The study of lenses constructed from fluids has a long-standing history. It has been observed that fluidic lenses can easily change their shape. Consequently, their focal length can be varied effortlessly, representing a key advantage of a liquid lens over conventional, solid lenses, whose curvature and focal length remain fixed. The given characteristic of a variable-focus fluidic lens is investigated in various papers \cite{optofluid, varfoc, varfocvolt, varfocservo, varfocachrom, liqmicro, varfocopt, varfoccam}. Several approaches for modulating the focal length of fluidic lenses are demonstrated, including the use of an external voltage \cite{varfocvolt} and a servo motor \cite{varfocservo}. Moreover, such variable-focus liquid lenses hold a robust practical potential for integration into photographic applications \cite{varfoccam}.

In addition to adjusting the focal length of the lens, the optical properties of a standard droplet have also been examined. Including a study of the magnification of a small water droplet emerging from a syringe \cite{projector}. By shining the laser beam through the droplet, the author obtained highly magnified projections of various microorganisms living in that water droplet. The author also provided a detailed derivation of the final magnification formula, mainly based on ray tracing and Snell's law. In a related study \cite{droplets}, the authors experimentally measured both the focal length and magnification of droplets and, notably, achieved considerable magnification of approximately 40x, enabling them to examine a diversity of biological tissues. Furthermore, additional research has been conducted on glass-placed water droplets. Starting with studies \cite{sb,mphotos} focusing on the magnifying effect of droplets, but they barely provide prominent analytical data. A more advanced paper \cite{lenseq} presents a straightforward mathematical approach with a classical lens equation and supplements it with experimental figures, underlining the significance of the volume of a droplet in terms of magnification. The same approach was used in another paper \cite{oil}, in which the author also examined various types of oil droplets, which are more resistant to evaporation, and compared their magnification to that of water droplets. Besides, it is notable that in the latter paper, the author calculated the resolution of an oil droplet employing the traditional USAF test target. 

In most of the above-mentioned articles, a droplet is considered a plano-convex lens, which, according to the paraxial approximation, converges rays at a single point where a resized/magnified image is formed. In addition to these assumptions, the shape of a droplet is considered spherical, which is valid for only small droplets (see Appendix~\ref{sphericalapprox}). In contrast, in our study, we primarily focus on larger, aspherical droplets. Therefore, we also investigate their shape. 

It is worth mentioning that these droplets are distinguished by their unique optical properties. In particular, likely to spherical lenses - but to a greater extent - the magnification is not uniform across the droplet that is deemed to be the main root of so-called pincushion and barrel distortions of the images observed in our experiments (see Fig.~\ref{pinc} and Fig.~\ref{barl}). Such distortions hinder a rigorous analysis of magnification, specifically, within a certain range of distance from the object, the images become so heavily distorted that accurate measurement of the dimensions of the magnified images becomes unfeasible. However, in this work, the already-mentioned distortions, or in other words, the variation of the magnification across a droplet, are examined.

Besides, an aspherical droplet is characterised by pronounced optical aberrations because, unlike spherical lenses, most paraxial rays do not converge at one point. This kind of aberration impinges upon the resolving power of the lens. Hence, our method of angular resolution calculation is based on the study of the aberration, which is related to the ray tracing technique.

This paper is organised in the following way. In Section~\ref{Theory}, we determine the shape of a droplet placed on a horizontal surface. In Section~\ref{Disc} we calculate theoretically both the magnification and resolving power of a droplet and discuss the correlation with the experimental results, and in Section~\ref{Summary} we summarise the study.

\section{\label{Theory}Theoretical Model} 
    \begin{figure}[htbp]
    \begin{subfigure}{.5\textwidth}
        \centering
        \includegraphics[width=\linewidth]{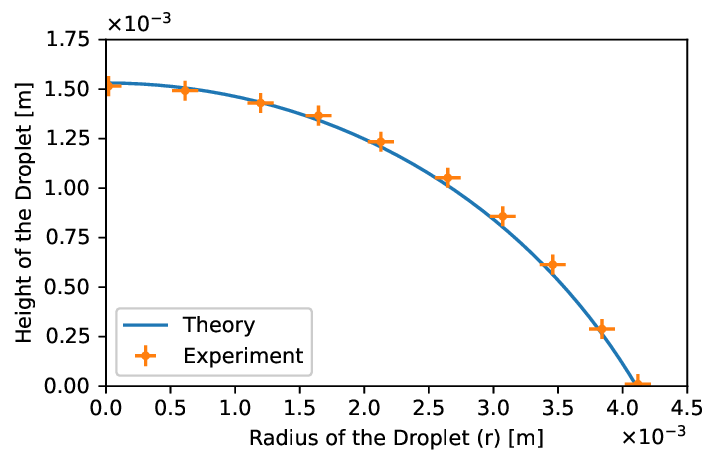}
        \caption{}
        \label{shape}
    \end{subfigure}%
    \begin{subfigure}{.4\textwidth}
        \centering
        \includegraphics[width=\linewidth]{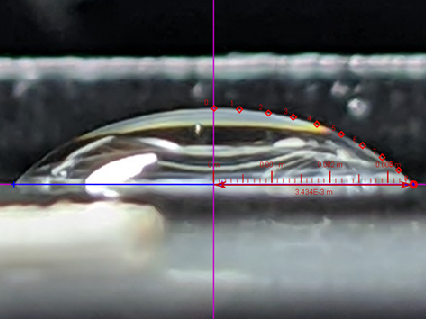}
        \caption{}
        \label{shape_setup2}
    \end{subfigure}
    \caption{(a) Comparison of the theoretical (line) and experimental (dots) shape of the droplet. The former is the numerical solution of  Eq.~\eqref{sys}, the system of differential equations. Experimentally, the shape of a droplet was determined by capturing a side-view image and fitting a set of boundary points between the droplet and the background. The limited resolution of the photo makes the boundary line not clearly visible, which leads to an error of several pixels. (b) A side-view photo of a droplet. The image was processed in Tracker (video, photo analysis and modelling tool), wherein the red points were marked along the droplet-background boundary line.}
    \end{figure}

    There are various ways to define the shape of a droplet on a flat surface. Although some papers use several approximations \cite{Vlado,Foroutan}, we describe it via the following system of equations \cite{parametrization,shape}:
    \begin{equation}
    \begin{cases}
        \frac{\rho g}{\sigma}f(r)-\frac{d\theta}{dl}-\frac{\sin{\theta}}{r}=c \\
        \frac{dr}{dl}=\cos{\theta}\\
        \frac{df}{dl}=\sin{\theta}
    \end{cases}
    \label{sys}
    \end{equation}
    where $\rho$ is the density of water, $f(r)$ is the function of the shape, $r$ is the radius of the droplet, $\sigma$ is the surface tension of water, $c$ is a constant, $dl$ is the differential length of the arc and $\theta$ is the angle between this arc and the horizontal axis.
    
    For a numerical solution of this system, we need three boundary conditions: $f(r_{max})=0$, $f'(0)=0$, $f'(r_{max})=-\tan\alpha$
    where $\alpha$ is the contact angle. 
    The solution of this system is plotted in Fig.~\ref{shape}, demonstrating excellent agreement with the experimental results. A way the latter was obtained is explicitly explained in Appendix~\ref{expsetup}, which also encapsulates a thorough description of all the measurement techniques and experimental setup.
    
\section{\label{Disc}Discussion}
In this section, we theoretically examine the magnification and angular resolution of a droplet and supplement it with experimental data.

\subsection{Magnification}
\begin{figure}[htbp]
    \centering
    \includegraphics[width=\linewidth]{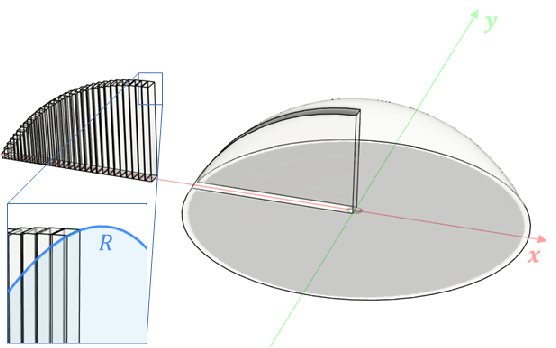}
    \caption{An image of a droplet (right) and an array of thick, spherical, plano-convex lenses, as a cut-out part of the droplet (left), each one curved along the $y$ axis. The lenses are symmetrically placed all around the droplet.}
    \label{lenses}
\end{figure}
In this subsection, we provide a comprehensive examination of the magnification of a droplet. 
We assume that a droplet contains an array of thick, spherical, plano-convex lenses, curved along the $y$ axis (Fig.~\ref{lenses}). It is worth mentioning that such an approximation enables us to calculate magnification across the droplet, as we can precisely calculate how the object under each lens will be magnified. However, it must be stated that this model is valid for the cases when the object is spread across the droplet (as in our experiments). In such cases, the following theoretical framework can describe how each part of the object will be magnified. 
Nevertheless, for instance, if a very small object is positioned at the centre of the droplet, only central lenses can accurately predict its magnified size, but the following theory will not work for the peripheral lenses because this object is far away from their optical axes. 
The radii of curvature of the lenses are defined with the following expressions:
\begin{equation}
    R=-\frac{\sqrt{1+\frac{\partial f(x,y)}{\partial y}^2}^3}{\frac{\partial^2 f(x,y)}{\partial y^2}}
    \label{R}
\end{equation}
where $f(x,y)=f(\sqrt{x^2+y^2})=f(r)$.

As for the magnification of each lens, we define it as the ratio of the distances between the droplet and the image ($q$) and the droplet and the object ($p$):
\begin{equation}
    M=\frac{q}{p}
    \label{m}
\end{equation}
The relation of $q$ and $p$ is described with the lens equation \cite{lens}:
\begin{equation}
    \frac{1}{p}+\frac{1}{q}=\frac{1}{f}
    \label{pq}
\end{equation}
and the focal length $f$ can be described with the lensmaker's equation \cite{lens}:
\begin{equation}
    \frac{1}{f}=(n-1)\left(\frac{1}{R}-\frac{1}{R'}+\frac{(n-1)b}{nRR'}\right)
    \label{f}
\end{equation}
where $n$ is the refractive index of the lens, $R$ and $R'$ are the radii of the convex and flat sides of the lens, respectively, and $b$ is the thickness of the lens. Since we assume that the lenses are plano-convex, the radius of the flat side ($R'$) equals infinity and Eq.~(\ref{f}) reduces to:
\begin{equation}
    \frac{1}{f}=\frac{n-1}{R}
    \label{f_final}
\end{equation}
Substituting Eqs.~(\ref{R}), (\ref{pq}) and (\ref{f_final}) into Eq.~(\ref{m}) the final formula of magnification is derived:


\begin{equation}
    M=-\frac{\sqrt{1+f'(x,y)^2}^3}{f''(x,y)\bigg|p(n-1)+\frac{\sqrt{1+f'(x,y)^2}^3}{f''(x,y)}\bigg|}
    \label{M_final}
\end{equation}

\begin{figure}[htbp]
\begin{subfigure}{.5\textwidth}
  \centering
  \includegraphics[width=\linewidth]{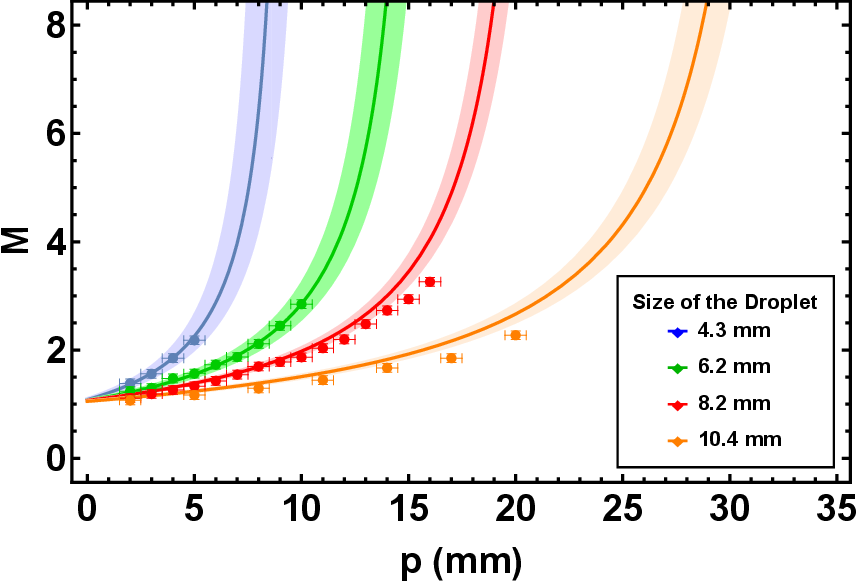}
  \caption{}
  \label{M_virt}
\end{subfigure}%
\begin{subfigure}{.5\textwidth}
  \centering
  \includegraphics[width=\linewidth]{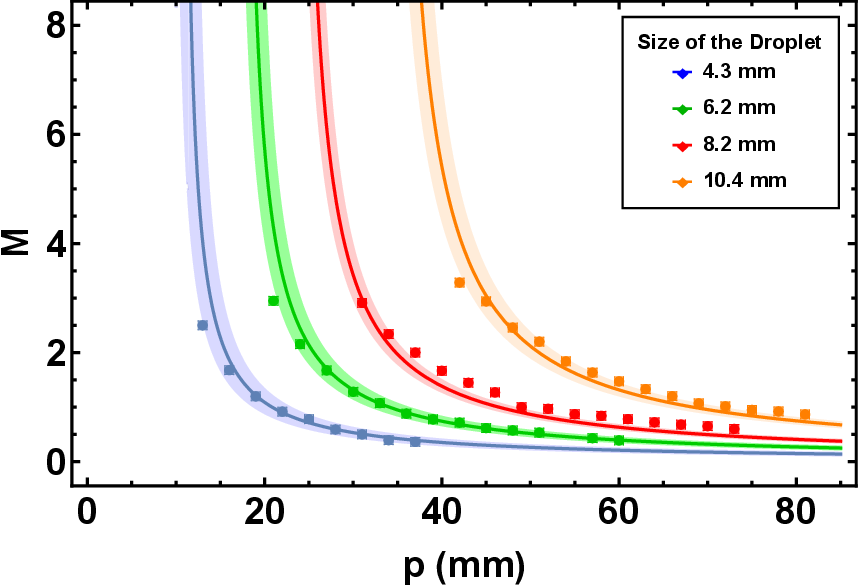}
  \caption{}
  \label{M_real}
\end{subfigure}
\begin{subfigure}{.5\textwidth}
  \centering
  \includegraphics[width=0.5\linewidth]{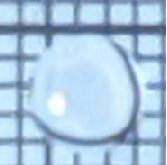}
  \caption{}
  \label{VHMR}
\end{subfigure}%
\begin{subfigure}{.5\textwidth}
  \centering
  \includegraphics[width=\linewidth]{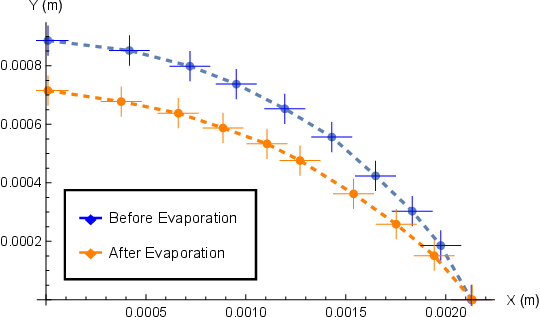}
  \caption{}
  \label{evap}
\end{subfigure}
\caption{Plots of theory (line) and experiment (dots) comparison of magnification of virtual image (a) and real image (b) depending on the distance between the object and the droplet when $r=0$. The theoretical errors are computed according to variations in the shape of the droplets because of evaporation. The magnification is theoretically calculated for a droplet before and after the evaporation, and the theoretical error, introduced as the difference between these results, is plotted as pale-colored areas along the solid lines, which represent the mean values. As for the experimental measurements of the magnification, the size of the magnified image was divided by the size of the object (see more in Appendix~\ref{expsetup}).  
(c) Image of a 4.3 mm size droplet, 10 mm from the object (graph paper). The image obtained in the droplet is distorted considerably.
(d) Shapes of a droplet before and after evaporation, which were obtained from the side-view images taken at the beginning and at the end of the magnification measurement process, respectively. To clarify, the photos were captured approximately 5 minutes apart.}
\label{M}
\end{figure}

Both Fig.~\ref{M_virt} and Fig.~\ref{M_real} show excellent theoretical and experimental correlation for all sizes of the droplets. It is evident that small droplets exhibit higher magnification in the virtual image region. However, when the image becomes real, the bigger droplets demonstrate superior magnifying capabilities. Furthermore, it is also worth mentioning that a considerable evaporation rate during the experiment led to the theoretical errors. More specifically, Fig.~\ref{evap} clearly illustrates that the entire shape of the droplet altered due to the evaporation, which caused the deviation in the derivatives included in the final magnification formula Eq.~(\ref{M_final}). Therefore, the magnification was calculated for both evaporated and non-evaporated/initial droplets, and the computation of the theoretical error was based on the difference between these results.

It is also apparent that the region where the droplet reaches maximal magnification (when the distance between the droplet and the object approaches the focal length of the central lens) is not experimentally examined because of the highly distorted images (Fig.~\ref{VHMR}).

\begin{figure}[htbp]
\begin{subfigure}{.5\textwidth}
    \centering
    \includegraphics[width=0.5\linewidth]{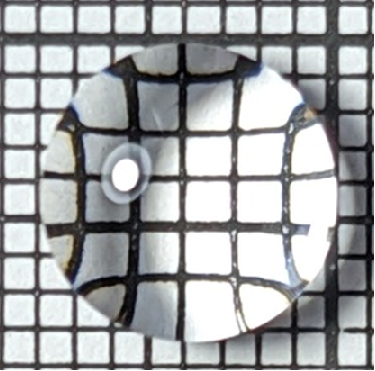}
    \caption{}
    \label{pinc}
\end{subfigure}%
\begin{subfigure}{.5\textwidth}
    \centering
    \includegraphics[width=0.5\linewidth]{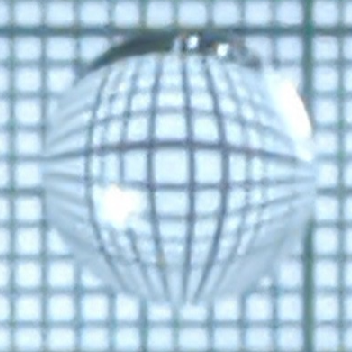}
    \caption{}
    \label{barl}
\end{subfigure}
\begin{subfigure}{.5\textwidth}
    \centering
    \includegraphics[width=0.9\linewidth]{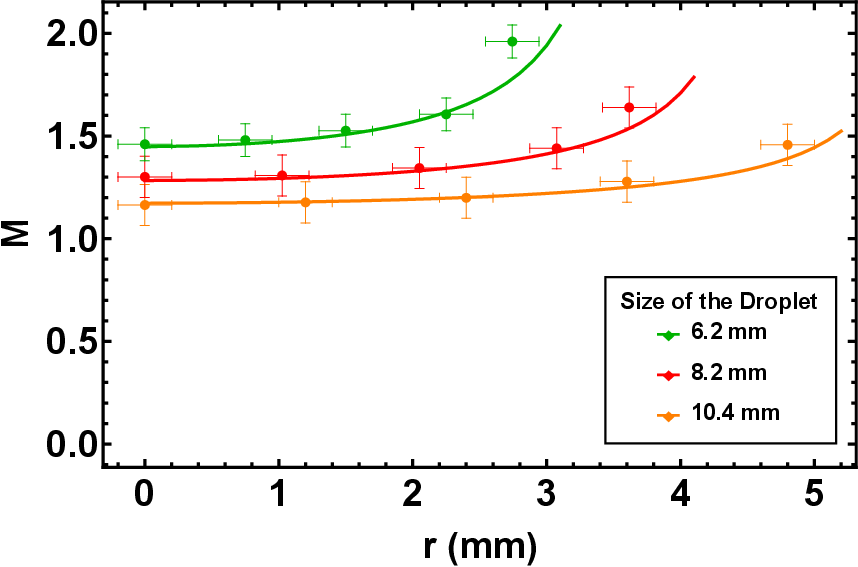}
    \caption{}
    \label{M_pinc}
\end{subfigure}%
\begin{subfigure}{.5\textwidth}
    \centering
    \includegraphics[width=0.9\linewidth]{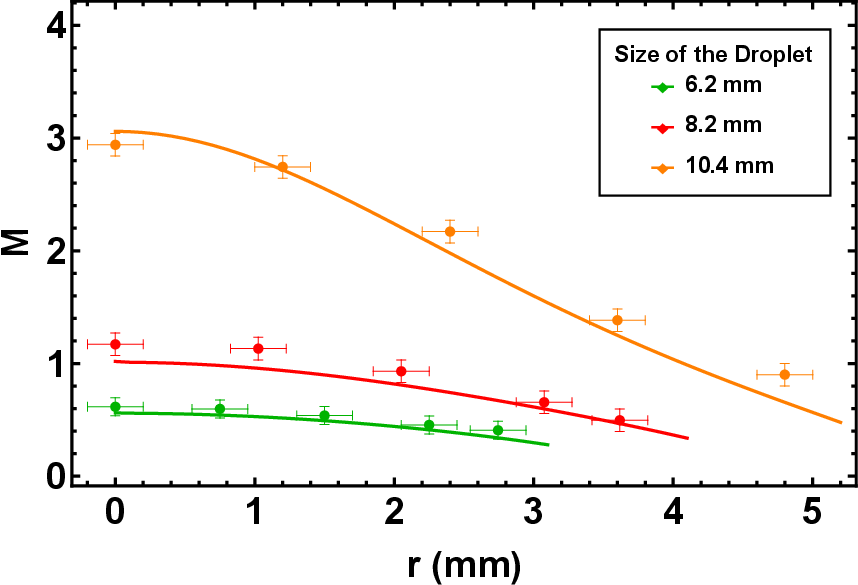}
    \caption{}
    \label{M_barl}
\end{subfigure}
\caption{(a) An image of the 8.2 mm size droplet, 5 mm away from the object (graph paper). (b) An image of the 8.2 mm size droplet, 46 mm away from the object (c) Comparison of the theoretical and experimental results of the virtual image magnification across the droplet (d) Comparison of the theoretical and experimental results of the real image magnification across the droplet.}
\label{M_dist}
\end{figure}

In addition to analysing the magnification of the central part of the droplet (i.e. at $r=0$), we also investigated the change of magnification across the droplet, the reason for the already-mentioned pincushion and barrel distortions of the virtual and real images, respectively. Fig.~\ref{M_pinc} and Fig.~\ref{M_barl} indicate that our theoretical framework can accurately predict the magnification changes for the droplets of varying dimensions. Interestingly, small droplets are characterised by a sharper magnification gradient of virtual images, whereas for real images, the gradient becomes steeper in larger droplets.

\subsection{Angular Resolution}
\begin{figure}[htbp]
    \centering
    \includegraphics[width=\linewidth]{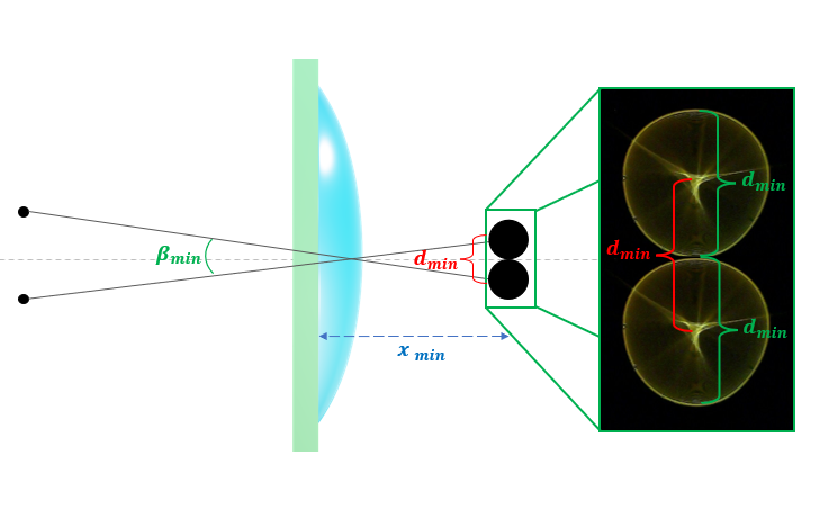}
    \caption{Two point-like objects (left side of the droplet) and their images (right side of the droplet), which are obtained on the image sensor and are seen as just separated. In our experiments, a laser emitting diverging rays was employed as the only light source, as it closely approximates a point-like object. Additionally, as both images have the same sizes and shapes, a single light source suffices to measure the minimal aberration spot diameter, which is essential for the calculation of angular resolution.
    }
    \label{resol}
\end{figure}

\begin{figure}[htbp]
\begin{subfigure}{0.5\textwidth}
    \centering
    \includegraphics[width=0.9\linewidth]{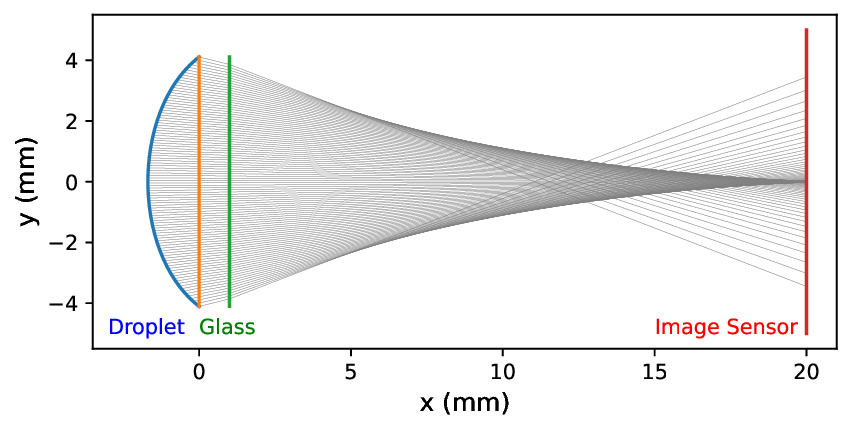}
    \caption{}
    \label{rays}
\end{subfigure}%
\begin{subfigure}{0.5\textwidth}
    \centering
    \includegraphics[width=0.9\linewidth]{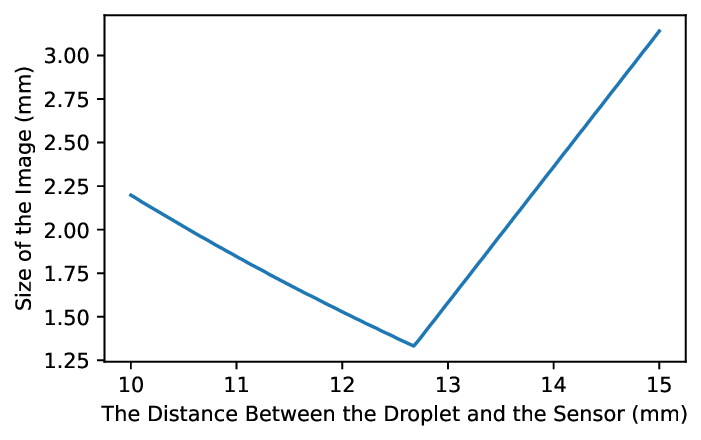}
    \caption{}
    \label{minsize}
\end{subfigure}
\caption{(a) Simulation of ray tracing (created in Python), where the point-like object is located at the left side 700 mm away from the droplet (the object and most refracted rays are omitted from the figure for clarity and visualization). (b) Result of the simulation. Size of the image/aberration spot as a function of the distance between the droplet and the image sensor (where the image is obtained)}
\label{sims}   
\end{figure}

In this subsection, the angular resolution of a droplet is analysed thoroughly.
To begin with, angular resolution depicts a minimal angle between two point-like objects that can be seen as just separated through a lens (Fig.~\ref{resol}). The primary reason for the limited angular resolution is that the images of these objects are not point-like, due to aberrations originating from the distinctive shape of the droplet. The angular resolution itself can be calculated with the following formula:
\begin{equation}
    \beta _{min}=\frac{d_{min}}{x_{min}}
\end{equation}
where $x_{min}$ is the distance between the images and the droplet, and $d_{min}$ is the distance between the two images when seen as just separated. It is worth noting that in the case of circular images, $d_{min}$ is equal to the minimal diameter of the image, in other words, the minimal aberration spot diameter. 
We use ray tracing (based on the vector form of Snell's law) to derive the final equation of the refracted ray:
\begin{equation}
    y_v=\frac{v_y}{v_x}(x-C_x)+C_y
    \label{refr_final}
\end{equation}
where all supplementary functions $v_y$, $v_x$, $C_x$, $C_y$ and detailed derivation of this equation are explicitly given in Appendix~\ref{raytracing}.
This equation is used to find the coordinates of the intersection point on the image sensor. The same technique is applied to thousands of rays (Fig.~\ref{rays}), and the image size is defined as the difference between the maximum and minimum $y$ coordinates of the intersection points. Subsequently, the image sensor is positioned at different locations along the optical axis, where the aberration spot diameters are calculated. This yields the resulting plot (Fig.~\ref{minsize}) and provides precise values of the two parameters of interest ($d_{min}$ and $x_{min}$).

\begin{figure}[htbp]
\begin{subfigure}{0.5\textwidth}
    \centering
    \includegraphics[width=0.9\linewidth]{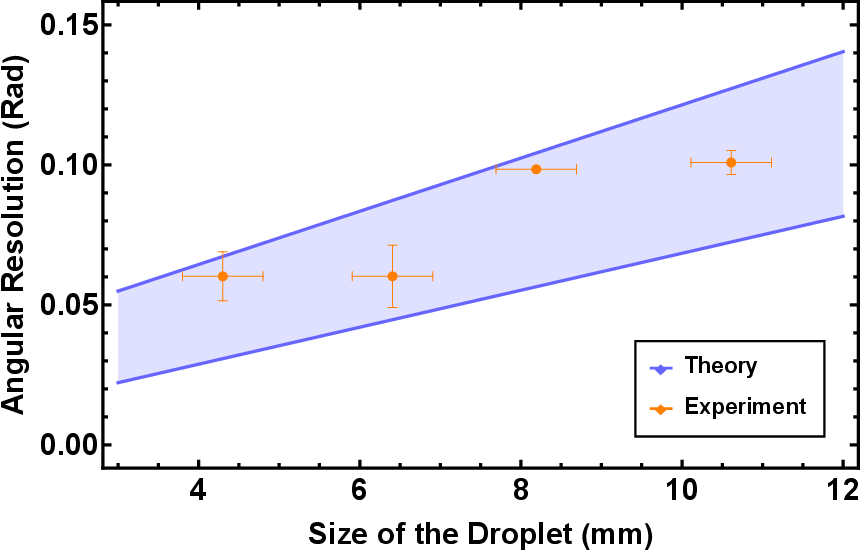}
    \caption{}
    \label{Ressize}
\end{subfigure}%
\begin{subfigure}{0.5\textwidth}
    \centering
    \includegraphics[width=0.9\linewidth]{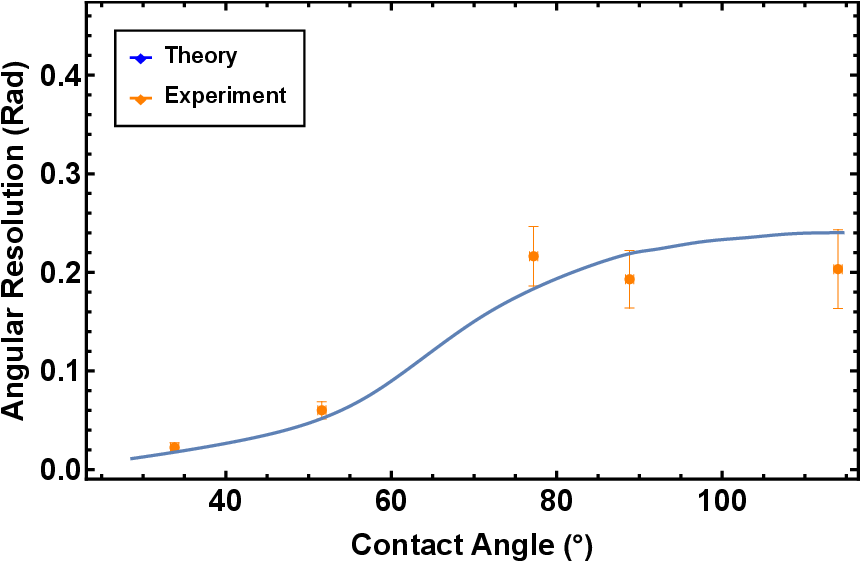}
    \caption{}
    \label{rescon}
\end{subfigure}  
\caption{(a) Plot of the angular resolution as a function of the droplet size. Since some droplets are not perfectly axisymmetric, the shape of the image/aberration spot is not perfectly circular, therefore, the sizes of these images vary along different axes, which leads to the uncertainty described via the vertical error bars.  As for the error bars on the horizontal axis, they show the standard measurement error of the ruler. (b) Plot of the angular resolution as a function of the contact angle. The errors are calculated similarly to the previous plot.}
\end{figure}

Fig.~\ref{Ressize} indicates the validity of our theoretical model. Notably, in this plot, the theoretical result is presented as a shaded region rather than a single curve because of the unpredictable contact angle hysteresis - i.e. the angle varies with the size of the droplet. Despite this fact, we plotted theoretical lines of the angular resolution versus the droplet size, for each contact angle obtained from the analysed droplets, and defined the region enclosed by these curves as the theoretical result.
Moreover, the plot reveals that big droplets exhibit poorer resolving power compared to smaller ones, due to increased asphericity, which is a key factor contributing to reduced angular resolution.

Additionally, Fig.~\ref{rescon} depicts that the resolving power of a water droplet is adversely affected by the increase of the contact angle, because of the analogous reason that hydrophobic droplets demonstrate enhanced aspherical properties. 

\section{\label{Summary}Summary}
For studying the magnification and the angular resolution, we examined the shape of the water droplet via the pressure balance and Young-Laplace equation. 
Next, we have assumed that the droplet contains an array of thick, spherical, plano-convex lenses with differing radii. 

The magnification of the central region of the droplet was examined using the lens and the lensmaker's equations. Nonetheless, the experimental study was constrained by significant image distortions when the distance between the object and the droplet approached the focal length of the central lens. 

Furthermore, taking into consideration that the lenses of the droplet have different radii, the magnification was also studied across the droplet, which is related to the study of the pincushion and barrel distortions. 

As for the angular resolution, we defined it as the ratio of the minimal aberration spot diameter and the distance between the latter and the droplet. Both parameters were computed using the ray tracing simulation, based on the vector form of Snell's law. 

Moreover, the influence of the size of the droplet on the magnification and the angular resolution was studied. It was revealed that small droplets exhibit superior magnifying and resolving capabilities than the bigger ones.

\section*{Acknowledgement}
The authors would like to thank T.Gachechiladze and G.Khomeriki for their support and the International Young Physicists' Tournament (IYPT) for fruitful discussions.

\section*{Disclosures}
The authors declare that there are no conflicts of interest related to this article.

\section*{Data Availability}
Data underlying the results presented in this paper are available in Ref.~\cite{Data}.

\section*{Appendix}

\setcounter{equation}{0}
\renewcommand{\theequation}{\thesection\arabic{equation}}

\setcounter{figure}{0}
\renewcommand{\thefigure}{\thesection\arabic{figure}}


\appendix

\section{\label{sphericalapprox}Validity of the Spherical Approximation}
\begin{figure}[htbp]
    \begin{subfigure}{0.5\textwidth}
        \centering
        \includegraphics[width=0.9\linewidth]{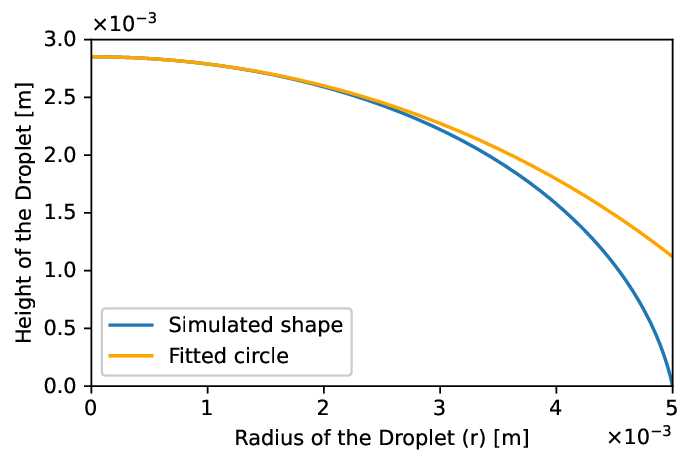}
        \caption{}
        \label{circlefit}
    \end{subfigure}%
    \begin{subfigure}{0.5\textwidth}
        \centering
        \includegraphics[width=0.9\linewidth]{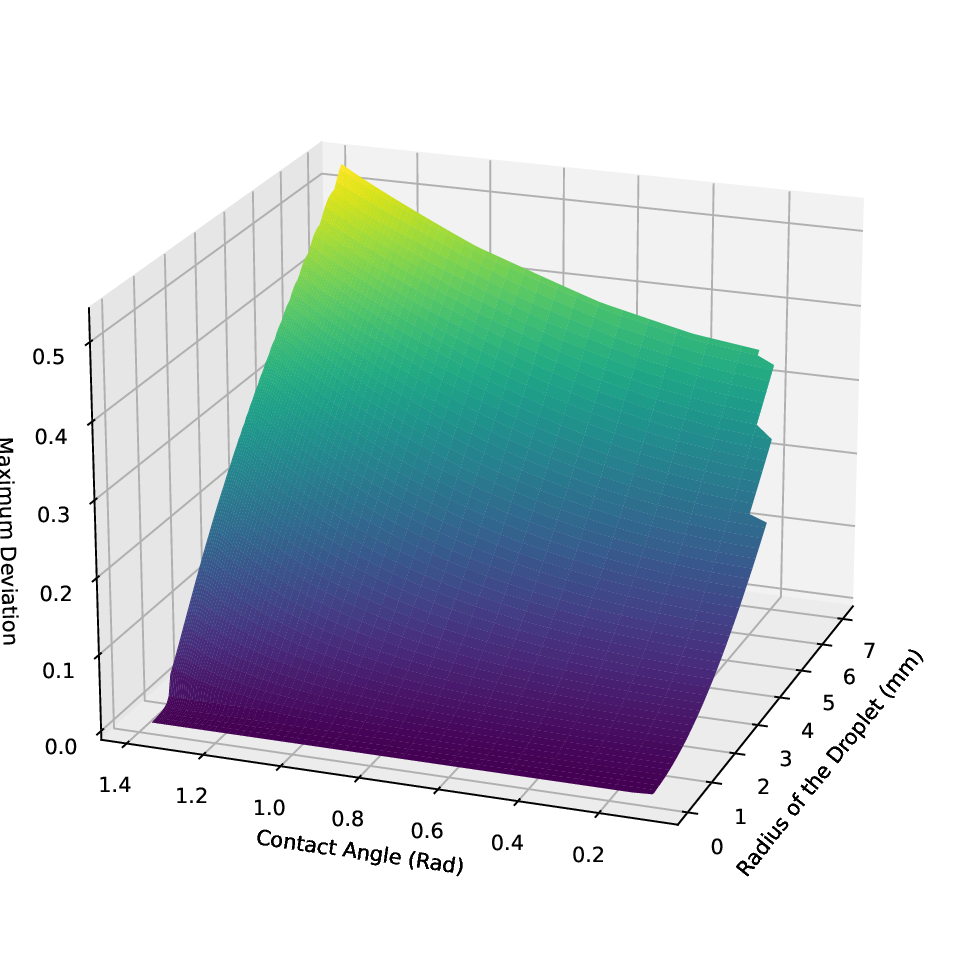}
        \caption{}
        \label{circledev}
    \end{subfigure}  
    \caption{(a) A plot of the simulated/theoretical shape of a droplet when $r_{max}=5\times10^{-3}m$ and $\alpha=1.249 
     Rad$ and the fitted circle, the radius of which equals the radius of the upper part of the droplet. (b) A 3D plot of the maximum deviation, a fraction of the maximum numerical difference between the simulated and spherical approximation results to the maximum height of the droplet, versus the radius of the droplet and the contact angle.}
\end{figure}

In this section, we evaluate the limits of the spherical approximation in terms of the shape of a droplet. Fig.~\ref{circlefit} indicates that the spherical approximation is not always valid. While the simulated shape aligns perfectly with the circular fit near the top of the droplet, a significant deviation arises toward the lower region. The circle is fitted in a way that at $r=0$ it coincides with the simulated shape $(f(r))$ and has the following radius:
 \begin{equation}
        R(0)=-\frac{1}{f''(0)}
\end{equation}
We introduced the maximal deviation as a fraction of the maximal difference between the circle fit $(y(r))$ and the simulation to the maximal height of the droplet:
\begin{equation}
    \epsilon=\frac{y(r_{max})-f(r_{max})}{f(0)}
\end{equation}
where $f(r_{max})=0$. Then we plotted the maximal deviation ($\epsilon$) versus the contact angle ($\alpha$) and the radius of the droplet ($r$). As Fig.~\ref{circledev} indicates, the smaller the latter two parameters, the smaller the deviation between the spherical approximation and the real shape.

\setcounter{equation}{0}
\setcounter{figure}{0}

\section{\label{raytracing}Ray Tracing}

\begin{figure}[htbp]
    \centering
    \includegraphics[width=\linewidth]{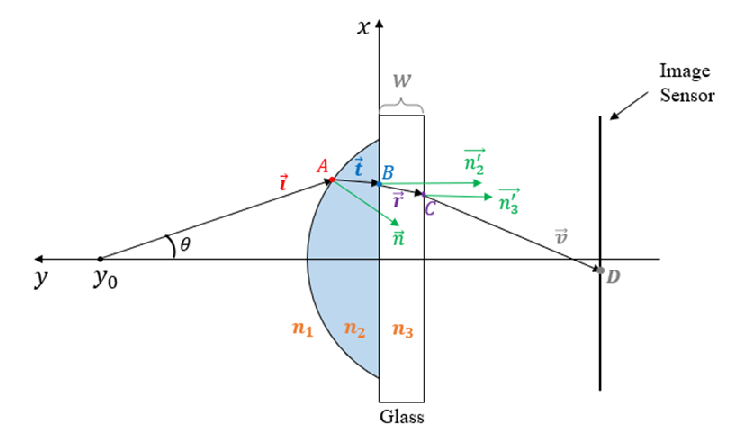}
    \caption{A detailed scheme of 2D ray tracing, where $n_1$, $n_2$ and $n_3$ are refractive indexes of air, water and glass respectively and $\overrightarrow{n}$, $\overrightarrow{n_2'}$, $\overrightarrow{n_3'}$ are normal vectors.}
    \label{oneray}
\end{figure} 

Below we give explicit expressions for all the supplementary functions used in Eq.~(16) and a detailed derivation of this equation.
Firstly, we calculate the coordinates of point A (Fig.~\ref{oneray}). For this we equalise the equations of the first ray $y=-x\cot\theta+y_0$ and the shape of the droplet $y=f(x)$ that gives us the coordinates of the point A:
\begin{equation}
    A_x=x, \quad
    A_y=-x\cot\theta+y_0
\end{equation}
    
where $y_0$ is the location of the light source and $\theta$ is the angle between the first ray and the y axis. Then, to write the equation of the refracted ray we use Snell's law in vector form:
\begin{equation}
    \overrightarrow{t}=\overrightarrow{n}\sqrt{1-n_1^2\frac{1-(\overrightarrow{n} \cdot \overrightarrow{i})^2}{n_2^2}}+\frac{n_1\overrightarrow{i}}{n_2}-\frac{n_1(\overrightarrow{n} \cdot \overrightarrow{i})\overrightarrow{n}}{n_2}
\end{equation}
where $n_1=1$ and $n_2=1.33$ are the refractive indexes of air and water respectively, $\overrightarrow{t}$ is refracted vector, $\overrightarrow{i}$ is incident vector components of which are ($\sin{\theta}$, $-\cos{\theta}$) and $\overrightarrow{n}$ is normal vector components of which are:
\begin{equation}
    n_x=\frac{f'(x)}{\sqrt{1+f'(x)^2}}, \quad 
    n_y=\frac{-1}{\sqrt{1+f'(x)^2}}
\end{equation}
Next, we introduce new variables such as $s$ and $d$ to simplify further calculations.
\begin{equation}
     s=n_x\sin{\theta}-n_y\cos{\theta}, \quad 
     d=\frac{\sqrt{n_2^2-(1-(n_x\sin{\theta}-n_y\cos{\theta})^2)}}{n_2}
\end{equation}
Finally we get the  components of $\overrightarrow{t}$:
\begin{equation}
    t_x=n_xd+\frac{i_x-sn_x}{n_2}, \quad t_y=n_yd+\frac{i_y-sn_y}{n_2}
\end{equation}
Equation of the $t$ ray $y_t=\frac{t_y}{t_x}(x-A_x)+A_y$ and the coordinates of the point B are as follows:
\begin{equation}
    B_x=-\frac{t_x}{t_y}A_y+A_x, \quad B_y=0
\end{equation}
After that, we use the same technique to write the equations of the rays $r$ and $v$:

\begin{eqnarray}
    r_x=\frac{n_2}{n_3}t_x, \quad r_y=-\frac{\sqrt{n_3^2-n_2^2(1-t_y^2)}}{n_3},\nonumber \\
    y_r=\frac{r_y}{r_x}(x-B_x), \nonumber \\
    C_x=B_x-\frac{r_x}{r_y}w, \quad C_y=-w, \nonumber \\
    v_x=n_3r_x, \quad v_y=-\sqrt{1-n_3^2(1-r_y^2)}, \nonumber \\
    y_v=\frac{v_y}{v_x}(x-C_x)+C_y
\end{eqnarray}
where $n_3=1.5$ is the refractive index of glass, $r_x$ and $r_y$ are the components of the vector $r$, $y_r$ is the equation of the  ray $r$, $C_x$ and $C_y$ are coordinates of point $C$, $w$ is the thickness of the glass, $v_x$ and $v_y$ are the components of the vector $v$ and $y_v$ is the equation of the  ray $v$.

\setcounter{equation}{0}
\setcounter{figure}{0}
\section{\label{expsetup}Experimental Setup}
\begin{figure}[H]
    \centering
    \includegraphics[width=0.5\linewidth]{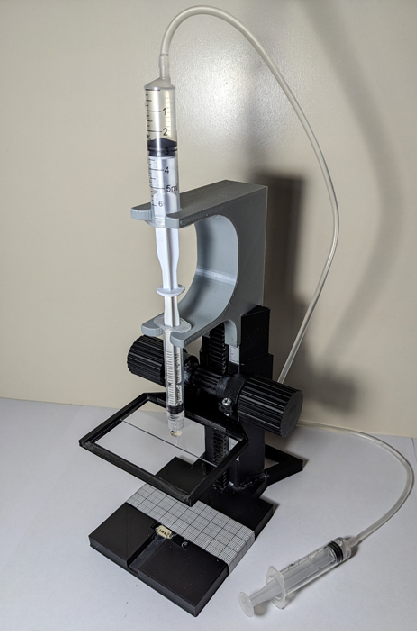}
    \caption{An image of the mechanism which was used to place a water droplet on a glass surface. It consists of a hydraulic system constructed using multiple syringes. By pressing the piston of the syringe, located in the bottom-right corner of the photo, the pushing force is transmitted via the liquid in the pipe, resulting in the displacement of the piston in the syringe from which the water droplet emerges.}
    \label{drop_setup}
\end{figure}

\begin{figure}
    \centering
    \includegraphics[width=0.5\linewidth]{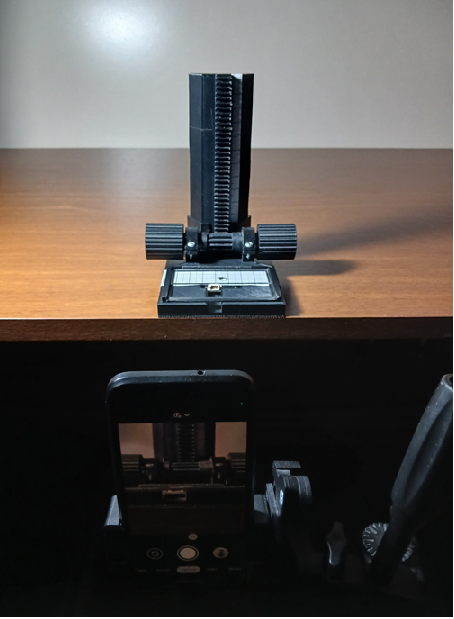}
   \caption{A photo of the experimental setup used for determining the shape of the droplet. A smartphone camera was used to capture the side-view photographs. The droplet's background colour and the light source orientation were adjusted to maximise image clarity.}
   \label{shape_setup}
\end{figure}

\begin{figure}[H]
    \centering
    \includegraphics[width=0.5\linewidth]{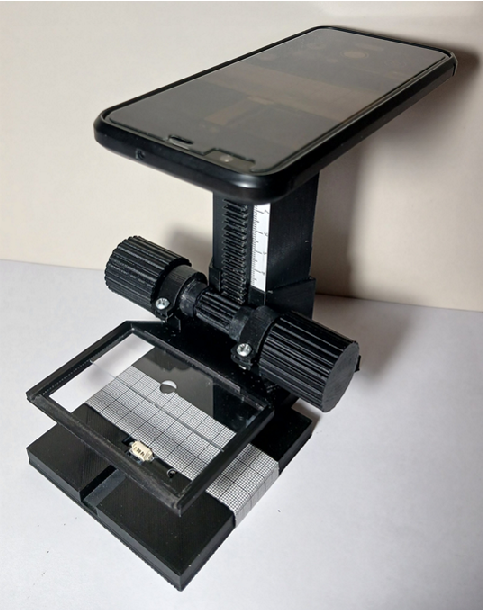}
    \caption{A photo of the mechanism utilised for measuring the magnification of the droplet. The 3D-printed rack and pinion actuator was useful in terms of varying the distance between the droplet and the object effortlessly. The latter in our case was a graph paper, and a top-view photo was captured using a manually-focused smartphone camera for measuring the magnified image size (the calibration was performed on the size of the droplet) and subsequently the magnification, by dividing the image size by the object size.}
    \label{M_setup}
\end{figure}

\begin{figure}[H]
\begin{subfigure}{0.5\textwidth}
    \centering
    \includegraphics[width=0.9\linewidth]{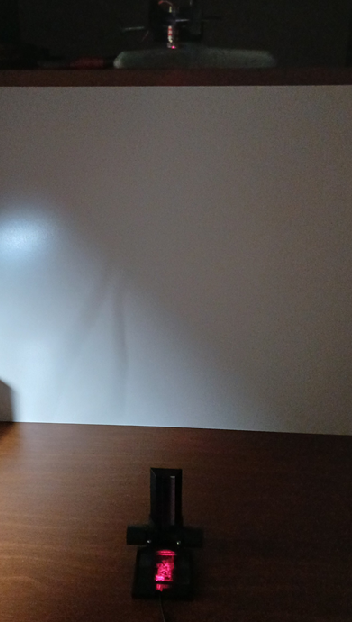}
    \caption{}
    \label{Res_set_1}
\end{subfigure}%
\begin{subfigure}{0.5\textwidth}
    \centering
    \includegraphics[width=0.9\linewidth]{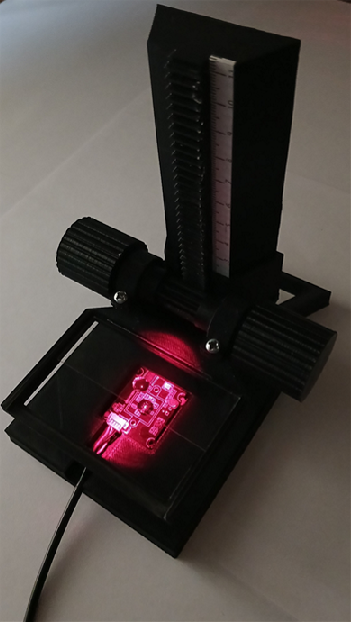}
    \caption{}
    \label{Res_set_2}
\end{subfigure}
\caption{(a) An image of the experimental setup used for measuring the angular resolution of the droplet. A 650nm laser, used as a point-like light source, was positioned 700 mm from the droplet with its beam diverged to illuminate the whole area of the droplet. (b) A photo of the mechanism used for measuring the angular resolution. The rack and pinion mechanism was used to change the distance between the droplet and the image sensor, where the image was obtained. The sensor was connected to a computer for the image analysis. }
\end{figure}

In this section, we provide a detailed explanation of the entire experimental process. Firstly, a water droplet was placed on a glass surface via the hydraulic mechanism shown in Fig.~\ref{drop_setup}. This mechanism played a crucial role in terms of avoiding direct contact with the syringe from which the water droplet emerges. Such an approach ensured that the droplet would be almost perfectly axisymmetric, as no hand oscillations would affect it. After positioning the droplet on the glass, the side-view image of the droplet was taken as depicted in Fig.~\ref{shape_setup}, which was then analysed (see Fig.~\ref{shape_setup2}) in order to experimentally determine the shape of the droplet. 
Afterwards, we started measuring the magnification using the mechanism, the image of which is given in Fig.~\ref{M_setup}. In addition to the fact that the rack and pinion mechanism greatly simplified the process of adjusting the distance between the droplet and the object, employing the graph paper as the object unfolded the opportunity to measure the magnification not only in the central part of the droplet but also across it. 
As for measuring the angular resolution, we used the setup which is given in Fig.~\ref{Res_set_1}. More precisely, in addition to the above-mentioned rack and pinion mechanism, we used a laser, which can be considered a point-like light source and an image sensor instead of a camera, to avoid the errors of the aberrations of built-in camera lenses.
After conducting experiments on a certain droplet, in most cases, we retook the side-view photos of the droplet in order to observe how the shape changed because of the evaporation. 

\end{document}